\documentstyle [preprint,aps,epsfig] {revtex}
\tightenlines
\begin{document}
\preprint{UFIFT-HEP-01-13}
\date{May 7, 2003} 
\title{Evidence for Ring Caustics in the Milky Way}

\author{Pierre Sikivie}
\address{Physics Department, University of Florida, Gainesville, FL 32611}

\maketitle

\begin{abstract}
The late infall of cold dark matter onto our galaxy produces discrete 
flows and caustics in its halo.  The recently discovered ring of stars 
near galactocentric distance 20 kpc and a series of sharp rises in
the Milky Way rotation curve are interpreted as due to the presence 
of caustic rings of dark matter in the galactic plane.  Their locations 
are consistent at the 3\% level with the predictions of the self-similar 
infall model for the caustic ring radii.  Also, a triangular feature in
the IRAS map of the galactic plane is consistent with the imprint of a 
caustic ring of dark matter upon the baryonic matter.  These observations
imply that the dark matter in our neighborhood is dominated by a single 
flow whose density and velocity vector are estimated.
\end{abstract}



There are compelling reasons to believe that the dark matter of the 
universe is constituted in large part of non-baryonic collisionless 
particles with very small primordial velocity dispersion, such as 
axions and/or weakly interacting massive particles (WIMPs).  Generically, 
such particles are called cold dark matter (CDM).  Knowledge of the 
distribution of CDM in galactic halos, and in our own halo in particular, 
is of paramount importance to understanding galactic structure and 
predicting signals in experimental searches for dark matter.

A model of the structure of the halos of isolated galaxies has been
developed \cite{ips,sty,cr} based on the observation that the dark 
matter particles must lie on a 3-dimensional sheet in phase-space, 
that this sheet cannot break, and hence that its evolution is 
constrained by topology.  The thickness of the sheet is the velocity 
dispersion.  The primordial velocity dispersion of the leading cold 
dark matter candidates is extremely small, of order
$10^{-12}c$ for WIMPs and $3\cdot 10^{-17}c$ (at most) for axions.  
To a coarse-grained observer, however, the sheet may have additional 
velocity dispersion because it is wrapped up on scales which are small 
compared to the galaxy as a whole.  This effective velocity dispersion 
is associated with the clumpiness of the dark matter before it falls 
onto the galaxy.  The effective velocity dispersion of the infalling 
dark matter must be much less than the rotation velocity of the galaxy 
for the model to have validity, say less than 30 km/s for our galaxy.  
On the other hand, by comparing the model with observations, an upper 
bound of order 50 m/s will be obtained below.

Where a galaxy forms, the sheet wraps up in phase-space, turning clockwise 
in any two dimensional cut $(x, \dot{x})$ of that space.  $x$ is the 
physical space coordinate in an arbitrary direction and $\dot{x}$ its
associated velocity.  The outcome of this process is a discrete set of
flows at any physical point in a galactic halo \cite{ips}.  Two flows 
are associated with particles falling through the galaxy for the first 
time ($n=1$), two other flows are associated with particles falling 
through the galaxy for the second time ($n=2$), and so on.  Scattering 
in the gravitational wells of inhomogeneities in the galaxy (e.g. 
molecular clouds and globular clusters) are ineffective in thermalizing 
the flows with low values of $n$.  Recently, Stiff and Widrow \cite{widr}
have put these discrete flows, also called 'velocity peaks', into evidence 
in $N$-body simulations using a new technique which increases the 
resolution of the simulations in the relevant regions of phase-space.

Caustics appear wherever the projection of the phase-space sheet onto 
physical space has a fold \cite{cr,sing,lens,Tre}.  Generically, caustics 
are surfaces in physical space.  On one side of the caustic surface
there are two more flows than on the other.  At the surface, the dark 
matter density is very large.  It diverges there in the limit of zero
velocity dispersion.  There are two types of caustics in the halos of 
galaxies, inner and outer.  The outer caustics are topological spheres 
surrounding the galaxy.  They are located near where a given outflow 
reaches its furthest distance from the galactic center before falling 
back in.  The inner caustics are rings \cite{cr}.  They are located 
near where the particles with the most angular momentum in a given 
inflow reach their distance of closest approach to the galactic center
before going back out.  A caustic ring is a closed tube whose 
cross-section is a $D_{-4}$ (also called {\it elliptic umbilic}) 
catastrophe \cite{sing}.  The existence of these caustics and their 
topological properties are independent of any assumptions of symmetry. 

Primordial peculiar velocities are expected to be the same for baryonic 
and dark matter particles because they are caused by gravitational forces.  
Later the velocities of baryons and CDM differ because baryons collide 
with each other whereas CDM is collisionless. However, because angular
momentum is conserved, the net angular momenta of the dark matter and 
baryonic components of a galaxy are aligned.  Since the caustic rings 
are located near where the particles with the most angular momentum in 
a given infall are at their closest approach to the galactic center, 
they lie close to the galactic plane.

A specific proposal has been made for the radii $a_n$ of caustic rings 
\cite{cr}:
\begin{equation}
\{a_n: n=1,2, ...\} \simeq (39,~19.5,~13,~10,~8,...) {\rm kpc} 
\times \left({j_{\rm max}\over 0.25}\right) \left({0.7\over h}\right) 
\left({v_{\rm rot} \over 220 {{\rm km} \over {\rm s}}} \right)
\label{crr}
\end{equation}
where $h$ is the present Hubble constant in units of 
$100\,{\rm km/(s~Mpc)}$, $v_{\rm rot}$ is the rotation velocity of the 
galaxy and $j_{\rm max}$ is a parameter with a specific value for each
halo.  For large $n$, $a_n \propto 1/n$.  Eq. \ref{crr} is predicted by 
the self-similar infall model \cite{ss,sty} of galactic halo formation.  
$j_{\rm max}$ is then the maximum of the dimensionless angular momentum 
$j$-distribution \cite{sty}.  The self-similar model depends upon a 
parameter $\epsilon$ \cite{ss}.  In CDM theories of large scale structure 
formation, $\epsilon$ is expected to be in the range 0.2 to 0.35 \cite{sty}.  
Eq. \ref{crr} is for $\epsilon = 0.3$.  However, in the range 
$0.2 < \epsilon < 0.35$, the ratios $a_n/a_1$ are almost 
independent of $\epsilon$.  When $j_{\rm max}$ values are quoted 
below, $\epsilon = 0.3$ and $h = 0.7$ will be assumed.

It was pointed out in ref. \cite{sty} that including angular momentum
in the self-similar infall model results in a depletion of the inner 
halo and hence an effective core radius.  The average amount of angular
momentum of the Milky Way halo was estimated \cite{sty} by requiring that
approximately half of the rotation velocity squared at our location is due
to dark matter, the other half being due to ordinary matter.  This yields 
$\bar{j} \sim 0.2$ where $\bar{j}$ is the average of the $j$-distribution
for our halo.  $\bar{j}$ and $j_{\rm max}$ are related if some assumption
is made about the shape of the $j$-distribution.  For example, if the 
$j$-distribution is taken to be that of a rigidly rotating sphere, 
one has $j_{\rm max} = {4 \over \pi} \bar{j}$. Hence 
$j_{\rm max} \sim 0.25$ for our halo.

Since caustic rings lie close to the galactic plane, they cause 
bumps in the rotation curve, at the locations of the rings.  In 
ref. \cite{kinn} a set of 32 extended well-measured rotation curves
was analyzed and statistical evidence was found for the $n=1$ and 
$n=2$ caustic rings, distributed according to Eq. \ref{crr}, at the
$2.6\sigma$ and $3\sigma$ significance level respectively.  That 
study implies that the $j_{\rm max}$ distribution is peaked near
0.27.  The rotation curve of NGC3198, one of the best measured, by 
itself shows three faint bumps which are consistent with Eq. \ref{crr} 
and $j_{\rm max} = 0.28$~\cite{cr}.   Also the above estimate of 
$j_{\rm max}$ for our own halo is close to the peak value of 0.27.

Eq. \ref{crr} with $j_{\rm max} = 0.25$ implies that our halo has 
caustic rings with radii near 40 kpc/$n$, where $n$ is an integer.
The purpose of this paper is to point out evidence in support of 
this extraordinary claim.  

Recently Yanny et al. \cite{yan} and Ibata et al. \cite{iba} 
discovered what appears to be a ring of stars, with radius of 
order 20 kpc, circling the Galaxy.  The origin of this star 
ring is puzzling.  In ref. \cite{yan}, the ring is interpreted 
as the tidal stream from an accreted satellite galaxy.  However, 
in this interpretation it is hard to account for the fact that 
the star population is confined to a nearly circular region 
\cite{iba}.  Also, it would have to be an accident that the 
tidal stream lies in the Galactic plane.  Ibata et al. \cite{iba} 
propose instead that the ring of stars is a perturbation of the 
Galactic disk population.  What causes the perturbation is not 
clear however.  Since the ring of stars is near the predicted 
$n=2$ caustic ring of dark matter, the perturbation may be the 
attractive gravitational field of the $n=2$ caustic ring.  This
interpretation accounts for the 20 kpc radius of the ring, as 
well as for the fact that the ring lies in the galactic plane.  

The spatial coincidence of the 20 kpc ring of stars with the 
predicted $n=2$ caustic ring may, of course, be fortuitous.  So 
it is natural to ask whether there is similar evidence for any of 
the other caustic rings.  The answer is yes for the $n=3$ ring. 
Binney and Dehnen studied \cite{bin} the outer rotation curve of 
the Milky Way and concluded that its anomalous behavior can be 
explained if most of the tracers of the rotation are concentrated 
in a ring of radius $1.6~r_\odot$ where $r_\odot$ is our distance to 
the galactic center.  Throughout this paper we use the standard value 
$r_\odot = 8.5$ kpc.  That value is assumed in Eq. \ref{crr}, and 
also in refs. \cite{yan,iba}.  The Binney and Dehnen ring is therefore 
at 13.6 kpc, which is within 3\% of the predicted radius of the $n=3$ 
caustic ring.  Moreover, there is independent evidence for the 
existence of the Binney and Dehnen ring.

\begin{figure}
\epsfxsize=6.5in
\centerline{\epsfbox{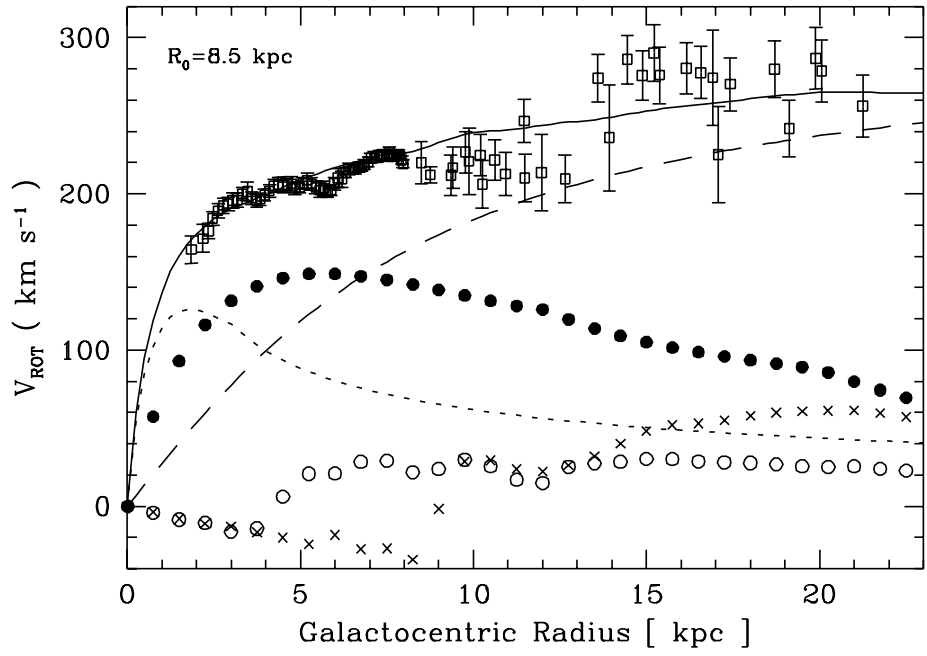}}
\vspace{-3.5cm}
\caption{Milky Way rotation curve from ref. [13].  The different lines 
represent the contributions from the bulge (dotted), the stellar disk
(filled circles), the HI layer (crosses), the H$_2$ layer (circles), 
and from a smooth dark halo (dashed).  The full line represents the 
sum of the contributions.  Reprinted by permission of the authors 
and Blackwell Publishing Ltd.}
\label{outrot}
\end{figure}

Olling and Merrifield have recently published \cite{olli} a rotation 
curve for the Milky Way.  It is reproduced in Fig. \ref{outrot}.  
It shows a significant rise between 12.7 and 13.7 kpc.  The increase 
in rotation velocity is 27\%, from 220 to 280 km/s.  A ring of matter 
in the Galactic plane produces a rise the rotation curve.  The rise 
expected from the $n=3$ caustic ring of dark matter, by itself, is 
only of order 3\%.  However, the effect of a caustic ring of dark 
matter is amplified by the ordinary matter (stars, gas, dust ..) 
which it attracts gravitationally.  The amplification would have to 
be by a factor of order nine in the case of the $n=3$ ring.  One may 
think, at first, that such as a large amplification is implausible
because the back reaction of the ring of ordinary matter upon the 
caustic ring of dark matter would determine the position of the 
latter, instead of the latter determining the position of the former.  
However this is not so because the dark matter caustic is not an 
overdensity of particles which are at rest with respect to the 
caustic ring.  The particles which at a given time make up the 
caustic ring are moving with great speed, of order 360 km/s for 
$n=3$, and are continually replaced by new particles.  As a result, 
the position of the caustic ring is insensitive to the gravitational 
field of the matter it attracts. 

The existence of rings of ordinary matter precisely where the $n=2$ 
and $n=3$ rings of dark matter are predicted may yet be fortuitous.  
Fig. 1 does not show a significant rise near the predicted location 
(10 kpc) of the $n=4$ caustic ring.  Note however that the error 
bars in Fig. 1 are very large for $r > r_\odot$.  The rise near 
10 kpc, if indeed there is one, may be too small to show up in 
the data.  On the other hand, the inner ($r < r_\odot$) part of 
the rotation curve is far better measured, and we may go look for 
rises there.

\begin{figure}
\epsfxsize=6.5in
\centerline{\epsfbox{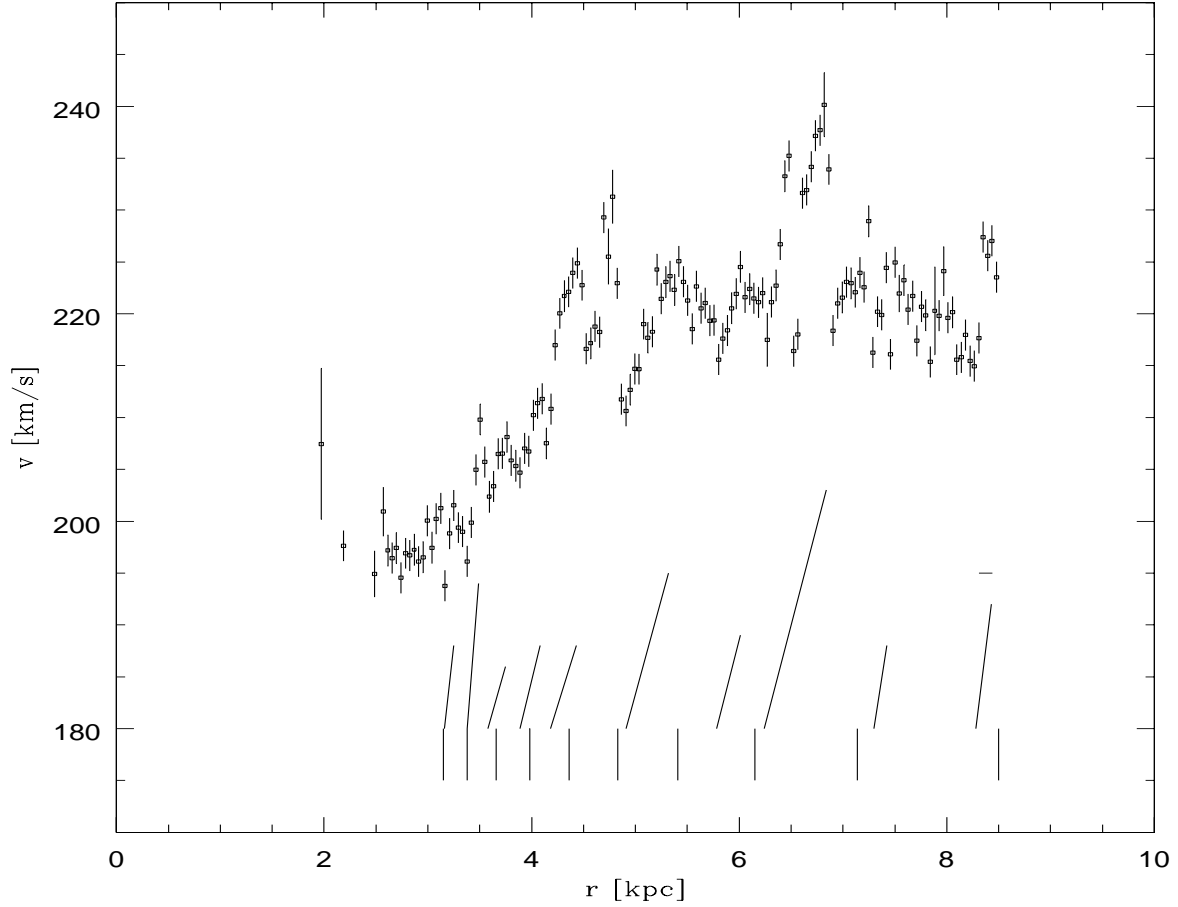}}
\vspace{1.5cm}
\caption{North Galactic rotation curve from ref. [15].  The locations
of the rises listed in the first three columns of Table I are indicated 
by line segments parallel to the rises but shifted downwards.  The 
caustic ring radii for the fit described in the text are shown as
vertical line segments.  The position of the triangular feature in 
the IRAS map of the galactic plane near $80^\circ$ longitude is shown 
by the short horizontal line segment.  It coincides with a rise in the 
rotation curve.} 
\label{inrot}
\end{figure}

Galactic rotation curves are obtained from HI and CO surveys of 
the Galactic plane.  A list of surveys performed to date is given 
in ref. \cite{merr}.  Everything else being equal, CO surveys 
have far better angular resolution than HI surveys because their
wavelength is nearly two orders of magnitude smaller (0.26 cm vs. 
21 cm).  The most detailed inner Galactic rotation curve appears 
to be that obtained \cite{clem} from the Massachusetts-Stony Brook 
North Galactic Plane CO survey \cite{CO}.  It is reproduced in
Fig. \ref{inrot}.  It shows highly significant rises between 3 
and 8.5 kpc.  Eq. \ref{crr} predicts ten caustic rings between 
3 and 8.5 kpc.  Allowing for ambiguities in identifying rises, 
the number of rises in the rotation curve between 3 and 8.5 kpc
is in fact approximately ten.  Below 3 kpc the predicted rises 
are so closely spaced that they are unlikely to be resolved in 
the data.

Table I lists ten rises, identified by the radius $r_1$ where they 
start, the radius $r_2$ where they end, and the increase $\Delta v$ 
in rotation velocity.  The rises are marked as slanted line segments 
in Fig. \ref{inrot}.  The evidence for the rise between 7.30 and 
7.42 kpc is relatively weak, so the corresponding numbers are 
in parenthesis in the Table.

The effect of a caustic ring in the plane of a galaxy upon its 
rotation curve was analyzed in ref. \cite{sing}.  The caustic 
ring produces a rise in the rotation curve which starts at 
$r_1 = a_n$, where $a_n$ is the caustic ring radius, and which 
ends a $r_2 = a_n + p_n$, where $p_n$ is the caustic ring 
width.  The ring widths depend in a complicated way on the 
velocity distribution of the infalling dark matter at last 
turnaround \cite{sing} and are not predicted by the model.  
They also need not be constant along the ring.

In the past, rises (or bumps) in galactic rotation curves have 
been interpreted as due to the presence of spiral arms \cite{spam}.  
Spiral arms may in fact cause some of the rises in rotation curves.
This does not, however, exclude the possibility of other valid 
explanations.  Two properties of the high resolution rotation 
curve of Fig. \ref{inrot} favor the interpretation that its 
rises are caused by caustic rings of dark matter.  First, there 
are of order ten rises in the range of radii covered (3 to 8.5 kpc).  
This agrees qualitatively with the predicted number of caustic rings,
whereas only three spiral arms are known in that range: Scutum, 
Sagittarius and Local. Second, the rises are sharp transitions 
in the rotation curve, both where they start $(r_1)$ and where 
they end $(r_2)$.  Sharp transitions are consistent with caustic 
rings because the latter have divergent density at $r_1 = a$ and 
$r_2 = a + p$ in the limit of vanishing velocity dispersion.  
Finally, there are bumps and rises in rotation curves measured 
at galactocentric distances much larger than the disk radius, 
where no spiral arms are seen.  In particular, the features 
found in the composite rotation curve constructed in ref. [9] 
are at distances 20 kpc and 40 kpc when scaled to our own galaxy.

The self-similar infall model prediction for the caustic ring radii, 
Eq. \ref{crr}, was fitted to the eight rises between 3 and 7 kpc by 
minimizing $rmsd \equiv [{1 \over 8} {\displaystyle \sum_{n=7}^{14}}
( 1 - {a_n \over r_{1 n}})^2]^{1 \over 2}$ with respect to $j_{\rm max}$, 
for $\epsilon = 0.30~$.  The fit yields $j_{\rm max} = 0.263$ and
$rmsd = 3.1\%~$. The fourth column of Table I shows the corresponding
caustic ring radii $a_n$.  In Fig. \ref{inrot} the latter are indicated 
by short vertical line segments. 

The velocity increase due to a caustic ring is given by 
\begin{equation}
\Delta v_n = v_{\rm rot} f_n {\Delta I(\zeta_n) 
\over \cos\delta_n(0) + \phi_n^\prime(0) \sin\delta_n(0)}\ .
\label{dv}
\end{equation}
The $f_n$, defined in ref. \cite{cr}, are predicted by the self-similar 
infall model, but $\Delta I(\zeta_n), \delta_n(0)$ and $\phi_n^\prime(0)$, 
defined in ref. \cite{sing}, are not.  Like the $p_n$, the latter 
parameters depend in a complicated way on the velocity distribution 
of the dark matter at last turnaround.  On the basis of the discussion 
in ref. \cite{sing}, the ratio on the RHS of Eq. \ref{dv} is expected 
to be of order one, but to vary from one caustic ring to the next.  The 
size of these fluctuations is easily a factor two, up or down.  The 
fifth column of Table I shows $\Delta v_n$ with the fluctuating ratio 
set equal to one, i.e. $\bar{\Delta} v_n \equiv f_n v_{\rm rot}$. 

For the reasons just stated, the fact that the observed $\Delta v$ 
fluctuate by a factor of order 2 from one rise to the next is 
consistent with the interpretation that the rises are due to 
caustic rings.  However the observed $\Delta v$ (column 3) are 
typically a factor 5 larger than the velocity increases expected 
from the caustic rings acting alone (column 5).  This is similar 
to what we found for the $n=3$ caustic ring, and suggests 
that the effects of the $n = 5 ... 14$ caustic rings are also 
amplified by baryonic matter they have accreted.  I'll argue 
that the gas in the disk has sufficiently high density and low 
velocity dispersion to produce such large amplification factors.  
I'll also give observational evidence in support of the hypothesis.

The equilibrium distribution of gas is:
\begin{equation}
d_{\rm gas} (\vec{r}) = d_{\rm gas}(\vec{r}_0) 
\exp[- {3 \over <v^2_{\rm gas}>}(\phi(\vec{r}) - \phi(\vec{r}_0))]\ ,
\label{gas}
\end{equation}
where $d$ is density and $\phi$ gravitational potential.  In the 
solar neighborhood, $d_{\rm gas} \simeq 
3\cdot 10^{-24} {{\rm gr} \over {\rm cm}^3}$ \cite{BT}, which is 
comparable to the density of dark matter inside the tubes of 
caustic rings near us.  From the scale height of the gas \cite{BT}
and the assumption that it is in equilibrium with itself and the 
other disk components, I estimate 
$\langle v_{\rm gas}^2 \rangle ^{1 \over 2} \simeq 8$ km/s.  The
variation in the gravitational potential due to a caustic ring over 
the size of the tube is of order $\Delta \phi_{\rm CR} \simeq 
2 f v_{\rm rot}^2 {p/a} \simeq (5 {{\rm km} \over {\rm s}})^2$.  
Because ${3 \over <v_{\rm gas}^2>} \Delta \phi_{\rm CR}$ is of order 
one, the caustic rings have a large effect on the distribution of 
gas in the disk.  The accreted gas amplifies and can dominate the 
effect of the caustic rings on the rotation curve.  To check
whether this hypothesis is consistent with the shape of the 
rises would require detailed modeling, as well as detailed 
knowledge on how the rotation curve is measured.  However, 
there is observational evidence in support of the hypothesis.

The accreted gas may reveal the location of caustic rings in maps of the
sky.  Looking tangentially to a ring caustic from a vantage point in the
plane of the ring, one may recognize the tricusp \cite{sing} shape of the
$D_{-4}$ catastrophe.  I searched for such features.  The IRAS map of the
galactic disk in the direction of galactic coordinates $(l,b) = (80^\circ,
0^\circ)$ shows a triangular shape which is strikingly reminiscent of the
cross-section of a ring caustic.  The relevant IRAS maps are posted at 
http://www.phys.ufl.edu/$\sim$sikivie/triangle/ .  They were downloaded
from the Skyview Virtual Observatory (http://skyview.gsfc.nasa.gov/).  
The vertices of the triangle are at $(l,b) = (83.5^\circ, 0.4^\circ),
(77.8^\circ, 3.4^\circ)$ and $(77.8^\circ, -2.6^\circ)$.   The shape 
is correctly oriented with respect to the galactic plane and the galactic
center.  To an extraordinary degree of accuracy it is an equilateral
triangle with axis of symmetry parallel to the galactic plane, as is
expected for a caustic ring whose transverse dimensions are small compared
to its radius.  Moreover its position is consistent with the position of a
rise in the rotation curve, the one between 8.28 and 8.43 kpc ($n=5$).  
The caustic ring radius implied by the image is 8.31 kpc and its 
dimensions are $p \sim 130$ pc and $q \sim 200$ pc, in the
directions parallel and perpendicular to the galactic plane
respectively.  It therefore predicts a rise which starts at 8.31 kpc 
and ends at 8.44 kpc, just where a rise is observed.  The probability 
that the coincidence in position of the triangular shape with a rise in 
the rotation curve is fortuitous is less than $10^{-3}$.

In principle, the feature at $(80^\circ, 0^\circ)$ should be matched 
by another in the opposite tangent direction to the nearby ring caustic, 
at approximately $(-80^\circ, 0^\circ)$.  Although there is a plausible 
feature there, it is much less compelling than the one in the
$(+80^\circ, 0^\circ)$ direction.  There are several reasons why 
it may not appear as strongly.  One is that the $(+80^\circ, 0^\circ)$ 
feature is in the middle of the Local spiral arm, whose stellar 
activity enhances the local gas and dust emissivity, whereas the 
$(-80^\circ, 0^\circ)$ feature is not so favorably located.  Another 
is that the ring caustic in the $(+80^\circ, 0^\circ)$ direction has
unusually small dimensions.  This may make it more visible by increasing 
its contrast with the background.  In the $(-80^\circ,0^\circ)$ direction, 
the nearby ring caustic may have larger transverse dimensions.

Our proximity to a caustic ring means that the corresponding flows, i.e. 
the flows in which the caustic occurs, contribute very importantly to the 
local dark matter density.  Using the results of refs. \cite{cr,sing,sty}, 
we can estimate their densities and velocity vectors.  Let us assume, for
illustrative purposes, that we are in the plane of the nearby caustic 
and that its outward cusp is 55 pc away from us, i.e. $a_5 + p_5 = 
8.445$ kpc.  The densities and velocity vectors on Earth of the $n=5$ 
flows are then:
\begin{equation}
d^+ = 1.7~10^{-24}~{{\rm gr} \over {\rm cm}^3}~,~
d^- = 1.5~10^{-25}~{{\rm gr} \over {\rm cm}^3}~,~
\vec{v}^\pm = (470~\hat{\phi} \pm ~100~\hat{r})~ 
{{\rm km} \over {\rm s}}, 
\label{lc}
\end{equation}
where $\hat{r}, \hat{\phi}$ and $\hat{z}$ are the local unit vectors
in galactocentric cylindrical coordinates.  $\hat{\phi}$ is in the 
direction of galactic rotation. The velocities are given in the 
(non-rotating) rest frame of the Galaxy.  Because of an ambiguity, 
it is not presently possible to say whether $d^\pm$ are the 
densities of the flows with velocity $\vec{v}^\pm$ or $\vec{v}^\mp$.
The large size of $d^+$ is due to our proximity to the outward cusp 
of the nearby caustic.  Its exact value is sensitive to our distance 
to the cusp.  We do not know that distance well enough to estimate 
$d^+$ with accuracy.  However we can say that $d^+$ is very large, 
of order the value given in Eq. \ref{lc}, perhaps even larger.  If we 
are inside the tube of the nearby caustic, there are two additional 
flows on Earth, aside from those given in Eq. \ref{lc}.  A list of 
local densities and velocity vectors for the $n \neq 5$ flows can 
be found in ref. \cite{bux}. An updated list is in preparation.

Eq. \ref{lc} has dramatic implications for dark matter searches.  
Previous estimates of the local dark matter density, based on 
isothermal halo profiles, range from 5 to 
7.5~$10^{-25}~{{\rm gr} \over {\rm cm}^3}$.  The present analysis 
implies that a single flow ($d^+$) has of order three times (or 
more) that much local density.

The sharpness of the rises in the rotation curve and of the triangular
feature in the IRAS map implies an upper limit on the velocity dispersion
$\delta v_{\rm DM}$ of the infalling dark matter.  Caustic ring
singularities are spread over a distance of order 
$\delta a \simeq {R~\delta v_{\rm DM} \over v}$ where $v$ is the velocity 
of the particles in the caustic, $\delta v_{\rm DM}$ is their velocity 
dispersion, and $R$ is their last turnaround radius.  The sharpness of 
the IRAS feature implies that its edges are spread over 
$\delta a \lesssim 20$ pc.  Assuming that the feature is due 
to the $n=5$ ring caustic, $R \simeq$ 180 kpc and $v \simeq 480$ km/s.  
Therefore $\delta v_{\rm DM} \lesssim 53$ m/s.  This bound is strongly 
at odds with the often quoted claim that structure formation on small 
scales causes all, or nearly all, late infalling dark matter to be in 
dwarf galaxy type clumps, with velocity dispersion of order 10 km/s.

The caustic ring model may explain the puzzling persistence of galactic
disk warps \cite{war}.  These may be due to outer caustic rings lying
somewhat outside the galactic plane and attracting visible matter.  
The resulting disk warps would not damp away, as is the case in 
more conventional explanations of the origin of the warps, but 
would persist on cosmological time scales.

The caustic ring model, and more specifically the prediction Eq.~\ref{lc}
of the locally dominant flow associated with the nearby ring, has 
important consequences for axion dark matter searches \cite{adm}, 
the annual modulation \cite{bux,ann,stiff} and signal anisotropy
\cite{anis,stiff} in WIMP searches, the search for $\gamma$-rays 
from dark matter annihilation \cite{gam}, and the search for 
gravitational lensing by dark matter caustics \cite{lens,mou}.  
The model makes predictions for each of these approaches to the 
dark matter problem.

The use of NASA's {\it Skyview} facility is gratefully acknowledged. 
I would like to thank Fred Hamann, Achim Kempf, Will Kinney, Heidi Newberg, 
Stuart Wick and the members of the Axion Dark Matter Search collaboration 
for valuable comments.  Finally I thank the Aspen Center for Physics for
its hospitality while this paper was being written.  This work is 
supported in part by U.S. DOE grant DEFG05-86ER-40272.


\begin{table}
\caption{Radii at which rises in the Milky Way rotation curve of Fig. 2
start ($r_1$) and end ($r_2$), the corresponding increases in velocity 
$\Delta v$, the caustic ring radii $a_n$ of the self-similar infall 
model for the fit described in the text, and typical velocity increases 
$\bar{\Delta} v_n$ predicted by the model without amplification due to
accretion of ordinary matter onto the caustic rings.} 
\vspace{0.5cm}
\begin{tabular}{ccccc}
$r_1$  &$r_2$&$\Delta v$ &    $a_n$        &$\bar{\Delta}v_n$\\
(kpc)  & (kpc) & (km/s) &     (kpc)        & (km/s)      \\
       &       &        &   n = 1 ..14     &  n = 1 .. 14  \\
\hline
       &       &       &      41.2         & 26.5  \\
       &       &       &      20.5         & 10.6  \\
       &       &       &      13.9         &  6.8  \\
       &       &       &      10.5         &  5.0  \\
8.28   & 8.43  &   12  &       8.50        &  3.9  \\
(7.30) &(7.42) &   (8) &       7.14        &  3.2  \\
6.24   & 6.84  &   23  &       6.15        &  2.6  \\
5.78   & 6.01  &    9  &       5.41        &  2.3  \\
4.91   & 5.32  &   15  &       4.83        &  2.0  \\
4.18   & 4.43  &    8  &       4.36        &  1.7  \\
3.89   & 4.08  &    8  &       3.98        &  1.5  \\
3.58   & 3.75  &    6  &       3.66        &  1.3  \\
3.38   & 3.49  &   14  &       3.38        &  1.2  \\
3.16   & 3.25  &    8  &       3.15        &  1.1  \\          
\hline
\end{tabular}
\label{tbl1}
\end{table}

\end{document}